# Surface Impedance Modeling of All-Dielectric Metasurfaces

Alessio Monti, *Senior Member, IEEE,* Andrea Alù, *Fellow, IEEE,* Alessandro Toscano, *Senior Member, IEEE,* Filiberto Bilotti, *Fellow, IEEE*

*Abstract*—We develop a simple and reliable analytical model that allows describing the electromagnetic response of all-dielectric metasurfaces consisting of a single layer array of high-permittivity spherical particles. By combining Mie theory with a bi-dimensional homogenization approach, we derive closed-form expressions of the electric and magnetic surface impedances exhibited by the metasurface and, thus, its reflection and transmission coefficients. The effectiveness of the proposed approach is validated through a set of full-wave simulations. The availability of the analytical model here developed allows a more in-depth understanding of the complex scattering response of these electromagnetic structures and enables the design of innovative devices operating throughout the whole electromagnetic spectrum, including, for example, unconventional reflectors for antennas, broadband optical mirrors, and highly efficient nanoantenna reflectarrays.

*Index Terms*—Metasurfaces; surface impedance; magnetic reflectors; nanostructured materials.

## I. INTRODUCTION

METASURFACES represent one the most popular fields of current research in artificial electromagnetic materials [1]-[4]. Similar to metamaterials, metasurfaces enable unconventional wave effects but through electrically thin, low-loss, inexpensive and feasible structures. The unique phenomena enabled by metasurfaces include but is not limited to perfect absorption [5], polarization transformation [6], wavefront shaping and beamforming [7]-[8], holography [9], asymmetric transmission [10], near-diffraction limit focusing [12], electromagnetic invisibility [12], etc.

The above-mentioned effects are usually obtained at microwaves and millimeter waves using structured surfaces consisting of one (or more) subwavelength metallic patterns printed onto a dielectric layer. However, the metallic parts unavoidably involve Ohmic losses affecting the performance of most metasurface-based devices. To overcome this issue, which is increasingly relevant at large frequencies, metasurfaces made only by dielectrics have been put forward and extensively investigated in the last years [13]-[31]. These structures – also referred to as *all-dielectric metasurfaces* - consist of bi-dimensional arrays of dielectric particles made of a high-permittivity material. When the size of the individual particle is comparable to the wavelength inside the dielectric, the particle itself exhibits both an electric and a magnetic induced dipole moment, whose spectral position and intensity can be engineered by acting on the particle shape and separation. All-dielectric metasurfaces are particularly popular at infrared and optical frequencies [13]-[22], where Ohmic losses of metals are significant, but they also enable important applications in the microwave range [26]-[27]. For example, water-based metasurfaces [30]-[31] have been the subject of extensive study in the past years because of their wide tunability characteristics. In addition, all-dielectric devices are particularly well suited for high power microwave applications, given their reduced Ohmic heating and the reduced formation of electric arcs at the field concentration points [28].

Despite the strong interest in all-dielectric metasurfaces, their design is still mainly based on numerical optimization processes [25]-[26] or complex analytical approaches [21]-[22]. The lack of a simple and reliable analytical model implies time-consuming design, which sometimes prevents a deep understanding of the involved physical phenomena. This especially applies to non-uniform metasurfaces [32], which are popular for applications related to the control of the beam shape and polarization. The availability of a simple boundary-condition model to be applied to each individual unit-cell would significantly reduce the efforts for designing non-periodic structures, as it has been proven for metallic metasurfaces (see, for instance, [33]). The aim of this contribution is to develop a simple and effective approach to homogenize all-dielectric metasurfaces made by spherical particles through effective electric and magnetic parameters. To do so, we combine Mie theory with a surface impedance homogenization approach, generalizing the one we have recently derived to describe the optical behavior of low-loss and lossy plasmonic metasurfaces [34]-[36], but now considering also induced magnetic polarizabilities and the spatial dispersion. This approach allows deriving a closed-form expression for the frequency-dispersive surface impedance exhibited by an array of high-index spherical

Manuscript received April 29, 2019. This work has been developed in the frame of the activities of the research contract MANTLES, funded by the Italian Ministry of Education, University and Research as a PRIN 2017 project (protocol number 2017BHFZKH).

A. Monti is with the Niccolò Cusano University, Rome, Italy (corresponding author, e-mail: alessio.monti@unicusano.it).
A. Alù is with CUNY Advanced Science Research Center, 85 St. Nicholas Terrace, 10031, New York, USA.
F. Bilotti and A. Toscano are with the Department of Engineering of ROMA TRE University, 00146 Rome, Italy.

particles and for its reflection and transmission coefficients. The analytical formulas are tested through a set of full-wave simulations that confirm their effectiveness even for very small inter-element separation distances. The availability of a simple analytical model allows deeply investigating and understanding the coupling effects that arise when Mie resonators are arranged in a lattice and predicting their overall electromagnetic response. In addition, we will show that the proposed approach can be used for designing several innovative devices. As relevant examples, we will discuss the design of unconventional microwave reflectors, broadband optical mirrors, and highly efficient reflectarray working in the visible spectrum.

The paper is organized as follows: in Section II, we will develop the analytical model needed to describe the electromagnetic behavior of all-dielectric metasurfaces composed by spherical particles. Section III is devoted to the numerical verification of the effectiveness of the proposed approach. Finally, Section IV describes how the proposed model can be employed for designing all-dielectric devices working at either microwave or optical frequencies.

## II. Surface Impedance of All-Dielectric Metasurfaces

The aim of this Section is to develop a theoretical model describing *analytically* the electromagnetic response (*i.e.,* reflection and transmission coefficients) of an all-dielectric metasurface. The homogenization technique we rely on is the coupled dipole method described, for instance, in [37], but it is here applied to a realistic geometry supporting both electric and magnetic dipole contributions. Some corrections to the original model, required to achieve reliable results in the case of non-deeply subwavelength particles, are introduced and commented.

A sketch of a typical all-dielectric metasurface is shown in Fig. 1(a). It consists of spherical particles made by a non-magnetic material with relative permittivity $\varepsilon_s$ arranged in an infinite square lattice. The array is embedded into a non-magnetic material with relative permittivity $\varepsilon_h$. Each particle has radius $a$ and is separated by the neighboring elements by a distance $d$. Both $a$ and $d$ are assumed to be small compared to the wavelength inside the host material at the frequency of interest. We also assume that the particles support both electric and magnetic induced dipole moments denoted with **p** and **m**, which are parallel to the impinging electric and magnetic field, respectively. The aim of this Section is to derive a homogenized representation of this class of metasurfaces for an external plane wave excitation, *i.e.,* to derive the values of some electromagnetic parameters that an infinitely thin homogenous sheet must have to exhibit the same electromagnetic behavior of the array. For brevity, we consider here the case of oblique TE illumination, *i.e.,* the electric field of the impinging wave is parallel to the *y*-axis. Since we are considering a single-layer array made by particles whose radius is small compared to the free-space operation wavelength and, thus, the overall structure is electrically thin, the most appropriate electromagnetic parameter to describe the metasurface behavior is the surface impedance [1]. Furthermore, given the spherical shape of the particles, their non-bianisotropic behavior and the sub-wavelength period that makes the anisotropy induced by the lattice negligible, we can assume that the response of the array is isotropic on the *x-y* plane and that its effective surface parameters are scalar.

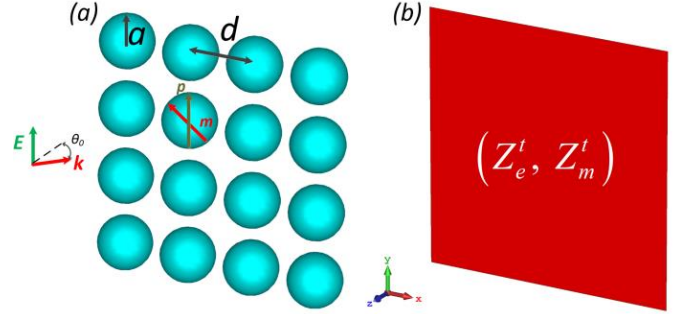

Fig. 1. (a) Sketch of an all-dielectric metasurface made by spherical particles illuminated by an external TE plane wave. (b) Homogenized representation of the all-dielectric metasurface.

The tangential electric $Z_e^t$ and the magnetic $Z_m^t$ surface impedances of the sheet link the total (incident + scattered) electric and magnetic fields to the electric and magnetic currents flowing on it through the following relations:

$$\mathbf{E} = Z_e^t \mathbf{J_s},$$
$$\mathbf{H} = \frac{1}{Z_m^t}\mathbf{M_s} = Y_m^t \mathbf{M_s}, \quad (1)$$

being $Y_m^t = 1/Z_m^t$ the effective tangential magnetic admittance of the metasurface.

The homogenization approach we rely on has been earlier described in [37]. According with it, the tangential electric surface impedance of an array of subwavelength particles separated by an electrically small distance can be written in the following compact form:

$$\frac{Z_e^t}{\eta_h} = -\frac{d^2}{k}\left[\left(\mathrm{Im}\left[\alpha_e^{-1}\right] - \frac{k^3}{6\pi}\right) + i\left(\mathrm{Re}\left[\alpha_e^{-1} - \beta_e\right]\right)\right], \quad (2)$$

where $k$ and $\eta_h$ are the wavenumber within and the intrinsic impedance of the host material, respectively, $\alpha_e$ is the electric polarizability of the individual scatterer, and $\beta_e$ is the interaction constant among the electric dipoles. In our scenario, however, the particles support an induced magnetic dipole moment, as well. Thus, by duality, we can write the tangential magnetic surface admittance of the array as follows:

$$\eta_h Y_m^t = -\frac{d^2}{k}\left[\left(\mathrm{Im}\left[\alpha_m^{-1}\right] - \frac{k^3}{6\pi}\right) + i\left(\mathrm{Re}\left[\alpha_m^{-1} - \beta_m\right]\right)\right], \quad (3)$$

where $\alpha_m$ and $\beta_m$ are the magnetic polarizability of the individual scatterer and the interaction constant among magnetic dipoles, respectively. It is worth noticing that, in the above expressions, we have assumed that the electric and magnetic dipoles do not interact each other. This is consistent with the discussion available in [38] for which only the parallel dipoles of the same nature (either electric or magnetic) do interact in a regular array.

To compute $Z_e^t$ and $Y_m^t$, we need a closed-form expression of the particle polarizabilities and of their interaction constants. As for the polarizabilities, in our earlier works, focused on the homogenization of electrically small plasmonic nanoparticles [34]-[36], we have employed the polarizability of the particle derived in static conditions. Here, this simple approach would lead to inaccurate results since the radius of the particles we are considering is generally comparable to the wavelength inside the dielectric material [17]. To overcome this issue, we need a more accurate description of the electromagnetic behavior of the single scatterer. This can be achieved by using the dynamic polarizabilities of the dielectric particle that read as [39],[40]:

$$\alpha_e = -i\left(\frac{6\pi}{k^3}\right)a_1, \quad \alpha_m = -i\left(\frac{6\pi}{k^3}\right)b_1, \quad (4)$$

where $a_1$ and $b_1$ are the dipolar Mie scattering coefficients of the individual particle. For the case of a sphere, the electric and magnetic Mie scattering coefficients can be written in the following compact form [41]:

$$a_n = \frac{n_s\psi_n(n_s x)\psi'_n(n_h x) - n_h\psi_n(n_h x)\psi'_n(n_s x)}{n_s\psi_n(n_s x)\xi'_n(n_h x) - n_h\xi_n(n_h x)\psi'_n(n_s x)},$$
$$b_n = \frac{n_h\psi_n(n_s x)\psi'_n(n_h x) - n_s\psi_n(n_h x)\psi'_n(n_s x)}{n_h\psi_n(n_s x)\xi'_n(n_h x) - n_s\xi_n(n_h x)\psi'_n(n_s x)}, \quad (5)$$

being $n_i = \sqrt{\varepsilon_i}$ (with $i = s, h$), $x = ka$, $\xi_n(...)$ and $\psi_n(...)$ the Riccati-Bessel functions and the prime denotes derivative with respect to the entire argument. It is worth noticing that, due to the radiation losses, the polarizabilities (4) are complex even in the ideal case of lossless scatterers. Specifically, it has been proven [42],[43] that the imaginary part of the inverse polarizabilities of a lossless scatterer are always equal to $-k^3/(6\pi)$. By introducing this result in (2)-(3), we obtain the following expressions for the surface impedance and for the magnetic admittance:

$$\frac{Z_e^t}{\eta_h} = -i\frac{d^2}{k}\text{Re}\left[\alpha_e^{-1} - \beta_e\right],$$
$$\eta_h Y_m^t = -i\frac{d^2}{k}\text{Re}\left[\alpha_m^{-1} - \beta_m\right]. \quad (6)$$

Thus, as expected, the surface impedance and admittance of an array of lossless dielectric particles is purely imaginary. It is also worth noticing that the surface impedance and admittance of our metasurface are expressed as the sum of two contributions: the first one is due to the scattering of the individual particle whereas the second term takes the coupling between the elements of the array into account.

To complete the model, we need to specify the expression of the interaction constant among nearby dipoles. Several efforts have been carried out during the past years for this purpose [44]-[47]. A comprehensive analysis of the interaction constant among electrically or magnetically polarizable particles can be found in a recent publication [38]. According to it, the interaction constants among parallel electric and magnetic dipoles in our scenario can be written as

$$\beta(\vartheta_0) = -\frac{i}{4d^2}\left[\begin{array}{c} 4k\left(\sin\frac{\vartheta_0}{2}\right)^4 + e^{ikR_0} \\ \left(\frac{2ik_tR_0J_0(k_tR_0) + kk_tR_0^2(\sin\vartheta_0)^2}{k_tR_0^2}\right) \\ \left(\frac{-2J_1(k_tR_0)(i+kR_0)}{k_tR_0^2}\right) \end{array}\right], \quad (7)$$

where $\theta_0$ is the angle between the impinging wavevector and the $z$ axis and $k_t = k\sin\vartheta_0$ is the transverse wavevector. In the case of normal incidence (i.e., $\theta_0 = 0°$), the interaction constant reduces to the well-known expression:

$$\beta_e = \beta_m = \frac{e^{ikR_0}}{4d^2R_0}(1+ikR_0) =$$
$$= \text{Re}\left[\frac{ik}{4d^2}\left(1+\frac{1}{ikR_0}\right)e^{ikR_0}\right] + i\left(-\frac{k^3}{6\pi} + \frac{k}{2d^2}\right), \quad (8)$$

being $R_0$ an approximate term equal to $R_0 = d/1.438$ (see, for instance, [37] – Section 4.5).

Equations (5)-(7) provide all the elements to derive, in closed-form, the expressions of the tangential electric surface impedance and of the tangential magnetic surface admittance exhibited by an all-dielectric metasurface when illuminated by an external TE plane wave. It is important to underline that, as can be appreciated in Fig. 1, a normal component of the magnetic dipole moment is also excited in the metasurface by the external field. However, as recently demonstrated [48], the presence of a normal polarization does not provide additional degree of freedom in engineering the metasurface response, as its effect can be described by equivalent tangential currents in the homogenization limit. In particular, a normal magnetic (electric) dipole radiates symmetrically (anti-symmetrically), similarly to a tangential electric (magnetic) dipole. Thus, the procedure described above for computing the tangential surface impedances can be extended to include also the normal magnetic surface impedance $Z_m^n$ as well.

Similarly, the described procedure can be applied for the case of TM incidence by using the corresponding expressions for the interaction constants in the TM scenario [38].

It is instructive to look at the analytical expressions we obtain in the case of a normally impinging plane wave. By inserting (8) into (2), we get

$$\frac{Z_e}{\eta_h} = i\left(\frac{d^2 \operatorname{Re}[\alpha_e^{-1}]}{k} - \frac{\sin(kR_0)}{4} + \frac{\cos(kR_0)}{4kR_0}\right)$$

$$\eta_h Y_m = i\left(\frac{d^2 \operatorname{Re}[\alpha_m^{-1}]}{k} - \frac{\sin(kR_0)}{4} + \frac{\cos(kR_0)}{4kR_0}\right), \quad (9)$$

where the real part of the inverse polarizabilities can be obtained combining (4) with (5). From the previous discussion, it should be clear that the first term of (9) is related to the response of the individual scatterer, whereas the other two terms account for the lattice contribution. After some analytical manipulations, we obtain the following expressions:

$$\operatorname{Re}[\alpha_e^{-1}] = \frac{k^3}{6\pi} \frac{A_e + B_e}{C_e + D_e},$$

$$\operatorname{Re}[\alpha_m^{-1}] = \frac{k^3}{6\pi} \frac{A_m + B_m}{C_m + D_m}, \quad (10)$$

with

$$A_e = y\cos(y)\left[\left(1 + (x^2 - 1)\varepsilon_s\right)\cos(x) - x(\varepsilon_s - 1)\sin(x)\right],$$

$$B_e = \sin(y)\left[(\varepsilon_s - 1)\cos(x) + x(\varepsilon_s - 1 + y^2)\sin(x)\right],$$

$$C_e = y\cos(y)\left[\left(1 + (x^2 - 1)\varepsilon_s\right)\sin(x) + x(\varepsilon_s - 1)\cos(x)\right],$$

$$D_e = \sin(y)\left[(\varepsilon_s - 1)\sin(x) - x(\varepsilon_s - 1 + y^2)\cos(x)\right], \quad (11)$$

and

$$A_m = x\varepsilon_s \sin(x)\sin(y),$$

$$B_m = \cos(x)(y\cos(y) + (\varepsilon_s - 1)\sin(y)),$$

$$C_m = -x\varepsilon_s \cos(x)\sin(y),$$

$$D_m = \sin(x)(y\cos(y) + (\varepsilon_s - 1)\sin(y)), \quad (12)$$

being $x = ka$ and $y = x\sqrt{\varepsilon_s}$. Equations (9)-(12) provide all the elements to write, in closed-form, the electric surface impedance and the magnetic surface admittance of an all-dielectric metasurface for normal incidence. It is worth noticing that all the three terms that concur to the surface impedance/admittance depend on the separation distance among the particles and, thus, on the coupling among the dipoles. In particular, the first term is proportional to $d^2$ whereas the dependency of the last two terms on d is contained in $R_0$. The second and the third terms, moreover, are oscillating functions with argument $kd$. This means that their contribution to $Z_e^t$ and $Y_m^t$ reaches its maximum at different frequencies.

Once the surface impedances are known, the reflection and transmission coefficients can be calculated as

$$\Gamma(\vartheta_0) = \frac{-\eta_0}{2Z_{symm}\cos(\vartheta_0) + \eta_0} + \frac{Z_{asymm}}{Z_{asymm} + 2\eta_0 \sec(\vartheta_0)},$$

$$T(\vartheta_0) = 1 - \frac{\eta_0}{2Z_{symm}\cos(\vartheta_0) + \eta_0} - \frac{Z_{asymm}}{Z_{asymm} + 2\eta_0 \sec(\vartheta_0)}, \quad (13)$$

being $\eta_0$ the free-space impedance, $Z_{symm} = Z_e^t + Z_m^n$ and $Z_{asymm} = Z_m^t$. When $\theta_0 = 0°$, the above formulas reduce to the well-known expressions available for a normally incident plane wave [49].

Before concluding this Section, we would like to emphasize two important aspects. First, the simple analytical model described above is quite general and can be easily extended to other geometries, such as coated spheres, anisotropic spheres and ellipsoids [41]. The only requirement, in fact, is the availability of a closed-form expression of the dipolar Mie scattering coefficients that is required to derive the dynamic polarizabilities (4). Analogously, by removing the assumption about the imaginary part of the polarizabilities we have used to derive (6), the model can be straightforwardly extended to lossy dielectric particles. Second, the surface impedances we have derived are valid even if the particles are not arranged in a regular lattice, such as the one shown in Fig. 1(a). In fact, the surface parameters (9) return an average description of the all-dielectric metasurface that does not depend critically on the exact positions of the particles in the array, as far as their size is small compared to the free-space wavelength.

### III. FULL-WAVE SIMULATIONS

In this Section, we check the effectiveness of the analytical model discussed above in several relevant scenarios through full-wave simulations. We first focus on the case of normal incidence and we analyze the electromagnetic behavior of three different all-dielectric metasurfaces made by spherical particles with refractive index $n_s = 4$ (i.e., $\varepsilon_s = 16$) embedded in air (i.e., $\varepsilon_h = 1$). The three metasurfaces differ for the radius of the particle that is equal to $0.08\lambda = 0.02\lambda_0$, $0.4\lambda = 0.1\lambda_0$ and $0.56\lambda = 0.14\lambda_0$, respectively, being $\lambda$ and $\lambda_0$ the wavelength inside the dielectric and the free-space wavelength at the frequency $f_0$. The inter-element separation distances between the elements of the metasurfaces are $0.09\lambda_0$, $0.3\lambda_0$ and $0.3\lambda_0$, respectively. It is worth noticing that, even if in some cases the particles have a radius comparable to the wavelength inside the dielectric, their size is still small compared to the free-space wavelength. This means that all the metasurfaces we are considering satisfy the conditions required to apply the proposed homogenization procedure.

In Fig. 2(a)-(c), we show the magnitude of the dipolar scattering coefficients $a_1$ and $b_1$ of the individual scatterer of the three metasurfaces calculated using (5). Given its subwavelength dimensions, the scattering response of the individual particle of the first metasurface is overall very small and is dominated by the electric dipole. As to the second case, a contribution of the magnetic dipole that dominates the scattering response above $f_0$ can be easily identified. Finally, the scatterer of the third metasurface, supports both an electric and a magnetic dipole moments that reach their maximum amplitude slightly above and below $f_0$, respectively.

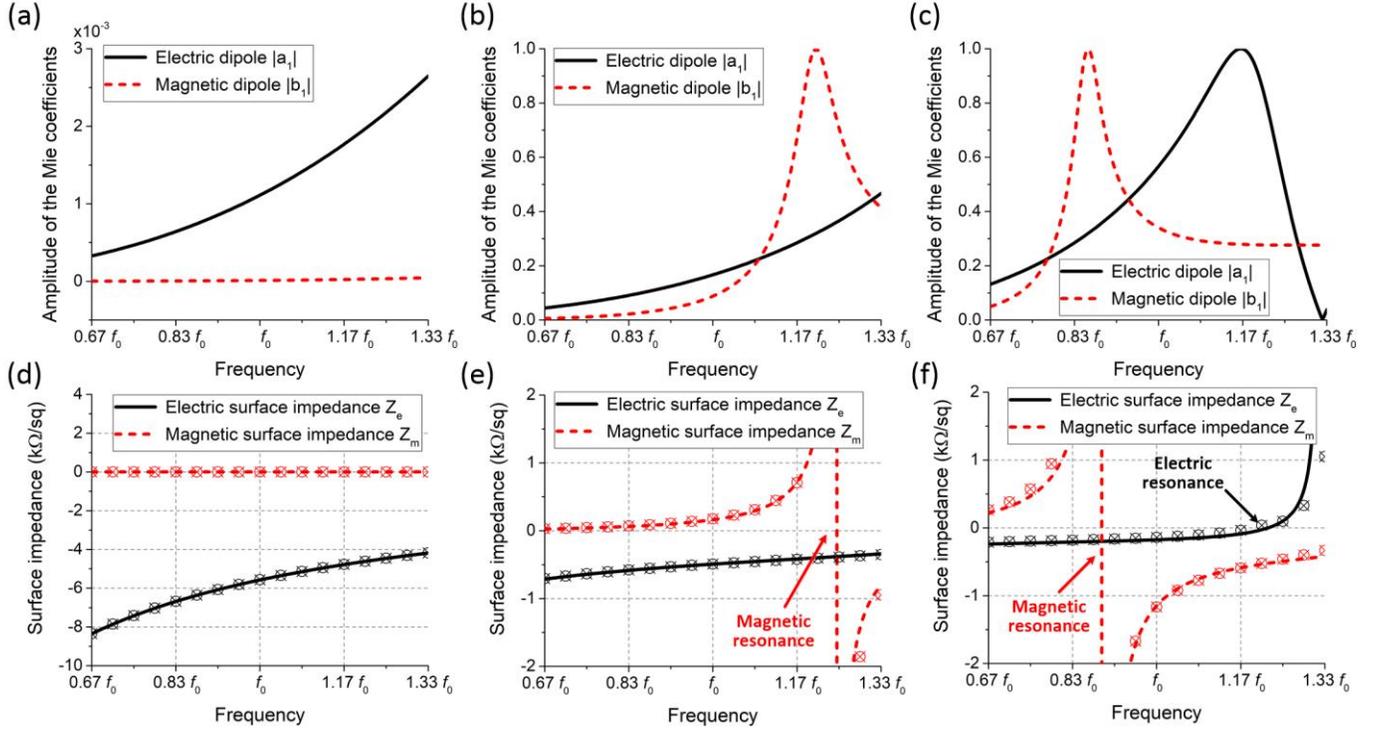

Fig. 2. Panels (a)-(c) depict the magnitude of the Mie scattering coefficients of the individual particles composing the three different all-dielectric metasurfaces described in the text. Panels (d)-(f) show the imaginary part of the effective surface impedances (both electric and magnetic) exhibited by the corresponding metasurfaces. Ticks represent the results retrieved through full-wave simulations. The geometrical parameters are $a = 0.02\lambda_0$ and $d = 0.09\lambda_0$, $a = 0.1\lambda_0$ and $d = 0.03\lambda_0$, $a = 0.14\lambda_0$ and $d = 0.03\lambda_0$ respectively.

The imaginary part of the surface impedances of the three corresponding metasurfaces are shown, respectively, in Fig. 2(d)-(f). We stress that the surface impedances describe the response of the whole all-dielectric metasurface, differently from the Mie scattering coefficients shown in panel (a)-(c) that refer to the electromagnetic response of the individual particle. In other terms, the surface impedance model complements the Mie analysis of the individual scatterer and allows predicting the response of an array of such particles. In these plots, the analytical surface impedances, calculated through (9)-(12), are compared to the ones retrieved through a numerical procedure based on full-wave simulations. Specifically, the procedure requires the numerical computation of the reflection and transmission coefficients of the metasurface unit-cell and the retrieval of $Z_e$ and $Z_m$ through the inversion of (13).

As a first consideration, we can observe an excellent agreement between theoretical and numerical results for all the considered scenarios. This confirms that the surface-impedance model we have described in the previous Section is able to predict accurately the electromagnetic response of the metasurface. Furthermore, these results also confirm that the surface impedance model catches the relevant physics behind all-dielectric metasurfaces. To better understand this point, let us consider the first scenario shown in Fig. 2(d). The model predicts a strong electric and a very low magnetic surface impedance. According to (1), this result indicates that the metasurface supports (low) electric currents, whereas the equivalent magnetic currents are vanishing. This is consistent with the analysis of the Mie scattering coefficients of the corresponding individual particle. Similar considerations can be carried out also for the second metasurface, whose response is dominated by the electric dipole before $f_0$ and, then, features a strong magnetic resonance. Finally, the third metasurface clearly exhibits two distinct resonances (i.e., $Z_m \to -i\infty$ and $Z_e = 0$) due to the magnetic and electric dipole moments supported by the Mie resonators.

To better appreciate the role of the coupling between electric and magnetic dipoles in an all-dielectric metasurface, in Fig. 3, we report the variation of the surface impedances as the inter-element separation distance $d$ changes. The metasurface considered here is composed by particles with radius $a = 0.572\lambda = 0.143\lambda_0$. As it can be appreciated, the inter-element separation distance has a strong effect on the average electric and magnetic impedance exhibited by the metasurface. This is an expected result and, as can be inferred from (9), it applies for any metasurface composed by Huygens meta-atoms. As it will be clearer in the next Section, such an effect can be exploited to engineer the overall response of the all-dielectric metasurface and achieve remarkable phenomena. The results shown in Fig. 3 also confirm that the proposed homogenization approach is effective for a wide range of inter-element separation distances and, remarkably, even if the particles are very close to each other (*i.e.*, $d$ very close to the diameter $2a$ of the particle).

Finally, we also check the validity of the model *vs.* the impinging angle $\theta_0$. As a relevant case, we investigate the spatial dispersion of the tangential magnetic surface impedance. We underline that this effect enables interesting features, as recently demonstrated [23]. We consider a metasurface made of dielectric particles with radius $a = 0.5\lambda = 0.082\lambda_0$ and separation distance $d = 0.2\lambda_0$. In Fig. 4, we show the magnetic surface impedance for three different incidence angles ($\theta_0 = 0°$, $\theta_0 = 30°$, $\theta_0 = 80°$, respectively). Also in this case, the theoretical results are compared to those retrieved by full-wave simulations. As expected, as the incidence angle increases, there is a red-shift of the magnetic resonance. The analytical model developed in this manuscript is able to well capture this phenomenon, even for large incidence angles, providing, thus, a powerful tool to estimate the frequency shift of the magnetic resonance that is relevant in several applicative scenarios.

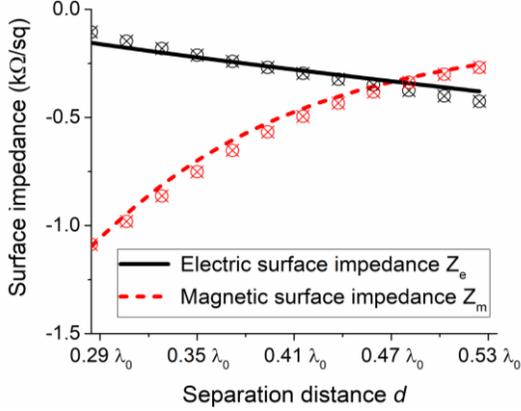

Fig. 3. Imaginary part of the surface impedances as a function of the inter-element separation distance of an array of dielectric particles with radius $a = 0.572\lambda = 0.143\lambda_0$. Ticks represent the results of the full-wave simulations.

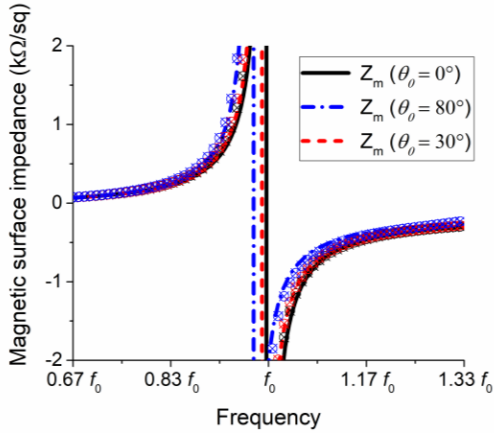

Fig. 4. Tangential magnetic surface impedance vs. the frequency for different values of the impinging angle $\theta$. The metasurface considered in this example has the following geometrical parameters: $a = 0.5\lambda = 0.082\lambda_0$ and $d = 0.2\lambda_0$. Ticks represent the results retrieved through full-wave simulations.

## IV. APPLICATIONS

In this Section, we apply the analytical model developed above to design devices with unconventional electromagnetic behavior. In all the cases considered below, the availability of an analytical model allows quickly achieving a final working design and better understanding the effect of the different parameters on the performance of the devices.

### A. *Optical reflectors*

As a first possible scenario where the surface impedance model of all-dielectric metasurfaces may play a role, we consider the design of optical reflectors. Conventionally, two possible approaches can be employed to design a device that is able to reflect all the impinging light within a desired range of frequencies. The first one is based on the use of dielectric materials. In particular, perfect reflection of light can be achieved by cascading periodically two quarter-wavelength layers made by different dielectrics. This device, also referred to as Bragg reflector, reaches high value of reflectivity within a frequency range whose width is determined by the ratio of the refractive indexes of the two dielectrics. Unfortunately, in order to achieve a reflectivity close to 100%, many layers are required and the overall thickness of the mirror is comparable (if not bigger) to the free-space wavelength [50]. The second conventional solution, instead, relies on the use of metallic coatings. In this latter case, broadband operation may be achieved but, due to the optical losses of metals, the coating reflectivity does not approach 100%. Clearly, both these solutions are not particularly suitable for modern nanophotonics where high performances and subwavelength devices are both required.

The idea we investigate here is to exploit the coupling between electric and magnetic dipoles supported by an all-dielectric metasurface to obtain a wide frequency range in which its reflectivity approaches 100%. In this case, the spherical particles are made by a high-index dielectric with low optical losses, i.e., the GaP, modeled through its complex measured permittivity function [51]. We set the radius of the nanoparticle to $a = \lambda/2$, being $\lambda$ the wavelength inside the dielectric at $f_0 = 450$ THz (i.e., in the red region of the optical spectrum, close to the working frequency of HeNe lasers). At $f_0$, the relative permittivity of GaP is $\varepsilon_s = 10.8 - i(0)$. The magnitude of the dipolar Mie scattering coefficients of the individual GaP spherical particle is shown in Fig. 5. As it can be appreciated, the two resonances (electric and magnetic) are quite close to each other.

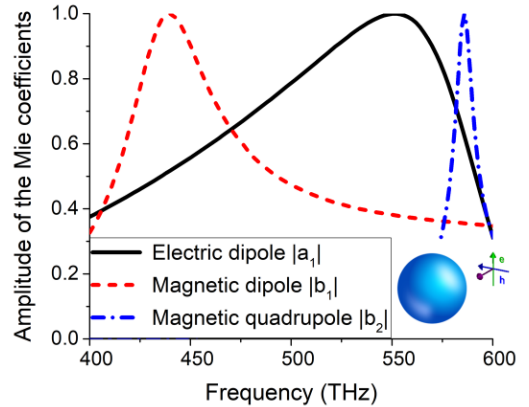

Fig. 5. Magnitude of the dipolar scattering coefficients and of the magnetic quadrupole of a GaP spherical particle with radius $r = \lambda/2$.

By exploiting the analytical model developed in the previous Section, we can easily calculate the reflectance $R$ (defined as $R = |\Gamma|^2$, being $\Gamma$ the reflection coefficient) of the metasurface for normal incidence and for different separation distance between the particles. The aim is to find an optimal value of $d$ to maximize the reflection bandwidth of the metasurface. The results (analytical and full-wave) are shown in Fig. 6 and confirm that, when the inter-element separation distance among the particles is equal to $1.28\lambda$ ($0.39\lambda_0$), the metasurface exhibits a single and wide reflection bandwidth. In particular, assuming a challenging threshold $R > 99\%$, a fractional bandwidth of 12.4% is achieved with a single and extremely thin layer ($t = 2a = 0.3\lambda_0$). The thickness of this all-dielectric reflector is significantly lower than the one of a conventional Bragg reflector based on a layered geometry and, thus, it can be easily integrated in modern nanophotonic devices. In addition, as it will be clearer in the next Section, the frequency phase response of these all-dielectric reflectors can be tailored to achieve unconventional phenomena.

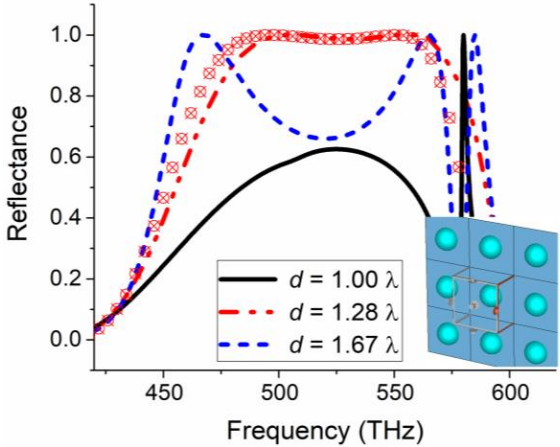

Fig. 6. Full-wave reflectance of an all-dielectric metasurface made by GaP spherical nanoparticles (radius $a = \lambda/2$) for different separation distances. The red ticked-curve represents the analytical results for the optimal case ($d = 1.28\ \lambda$). The inset shows the sketch of the all-dielectric metasurface.

We would like emphasizing that, for lower separation distances, the reflection coefficient does not reach high values. Conversely, if we increase the separation distance above the optimal value, the broadband reflection behavior tends splitting in a dual-band one. This behavior can be explained by looking at the analytical expression describing the variation of the surface impedances *vs.* the inter-element separation distance reported in Fig. 7 and Fig. 8. When the separation distance is low, the electric and magnetic resonances exhibit a reduced quality factor, i.e., the electric (magnetic) surface impedance tends to zero ($-i\infty$) in the frequency region between the two resonances. When this happens, the reflection coefficient (13) tends to zero, due to the balanced electric and magnetic response of the array. In contrast, increasing the separation distance between the dipoles results in increased quality factors of the resonances, i.e., the electric (magnetic) surface impedance deviates from (moves close to) zero in the intermediate region (green arrow). This causes the splitting of the broadband reflection behavior into a dual-band one.

Thus, an optimal value of the separation distance for which the quality factor of the resonances results in a single and wide reflection bandwidth must exist. Another interesting observation, arising from the observation of the analytical results, is related to the red shift of the electric and magnetic resonances as the separation distance increases (orange arrow). These results about optical reflectors are consistent with the ones discussed in [18] and are here supported by a rigorous formulation.

Before concluding this Subsection, we would like to mention that, in the full-wave results shown in Fig. 6, an additional narrow reflection peak appears slightly above the reflection bandwidth. This is due to the magnetic quadrupole of the individual sphere (shown in Fig. 5) that is omitted in our dipolar model. In spherical particles, however, the magnetic quadrupole resonance always appears at higher frequencies compared to the electric dipolar resonance. Thus, its presence does not affect the effectiveness of the surface impedance model within the reflection range.

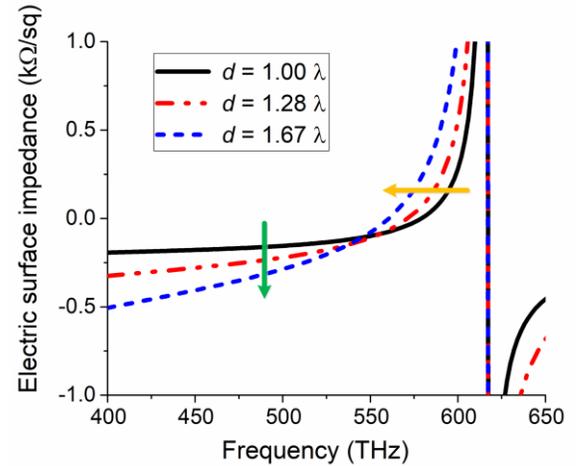

Fig. 7. Imaginary part of the electric surface impedance of the all-dielectric metasurface made by GaP spherical nanoparticles for different separation distances.

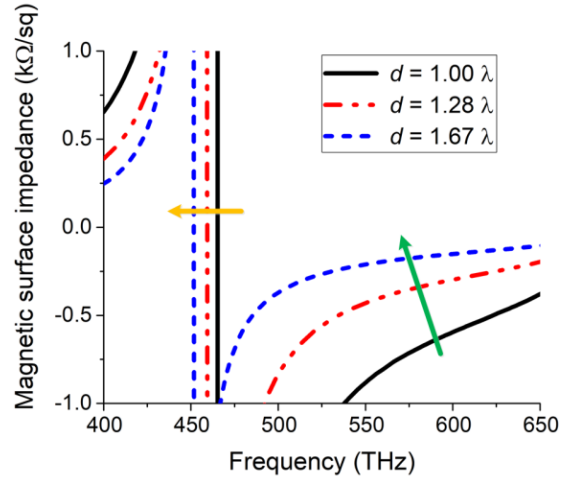

Fig. 8. Imaginary part of the magnetic surface impedance of the all-dielectric metasurface made by GaP spherical nanoparticles for different separation distances.

## B. Highly-efficient reflectarrays based on dielectric resonator antennas

Another application where the surface impedance model of an all-dielectric metasurface may play an important role in the design of optical dielectric resonator antennas (DRAs). Though there exists alternative devices, such as meta-gratings [52],[53], which allow achieving efficient extreme beam manipulation, reflectarrays are still the most convenient when moderate deflection angles are required. In the last years, several efforts have been carried out in this direction because optical DRAs may, in principle, overcome the losses limitations affecting plasmonic nanoantennas [54],[56]. However, the current configurations of optical DRAs still require a plasmonic element, i.e., the metallic ground plane that is needed to achieve full reflection of the impinging light, which limits the maximum achievable efficiency.

As an example, in [54], authors have discussed the design of an optical reflectarray based on DRAs and consisting of several $TiO_2$ cylindrical resonators put onto a silver layer. As discussed by the authors, the maximum efficiency of this device is around 70% and is limited by both metal losses and the excitation of a surface plasmon polariton at the interface between silver and air. Here, we show that the efficiency of a DRA-based optical reflectarray can be significantly increased by replacing the lossy plasmonic ground plane with an all-dielectric metasurface.

The structure we considered is shown in Fig. 9 and consists of three layers. The top layer is an array of seven $TiO_2$ cylindrical resonators with different diameters. The cylinders have been modeled using an anisotropic relative permittivity with values equal to 8.29 and 6.71 along the planar and the cylindrical axes, respectively, and by considering a loss tangent equal to 0.01. The lowest layer of the reflectarray, instead, is an all-dielectric metasurface that behaves as a perfect reflector at the desired frequency of operation. The middle layer is a silica spacer with thickness $h = 500$ $nm$ whose aim is to reduce the coupling between the metasurface and the array of dielectric resonators placed above. The all-dielectric backing reflector has been designed using the surface impedance model described in Section II. Its geometrical values are $a = 120$ $nm$ and $d = 310$ $nm$, corresponding to a reflection coefficient equal to $1\angle 147°$ at the working frequency $f_0$. The period of the reflectarray ($p = 310$ $nm$), its operation frequency ($f_0 = 475$ THz) and the other relevant parameters (such as the desired deflection angle $\theta = 20°$) have been chosen as in [54].

The optical response of the unit-cell of the reflectarray for a normally impinging plane-wave is shown in Fig. 10 as a function of the diameter of the cylindrical resonator. The reflection coefficient phase covers a quite large angular range (around 300°), while the magnitude is always close to 1. The overall optical losses are very low, due to the absence of plasmonic materials in the design.

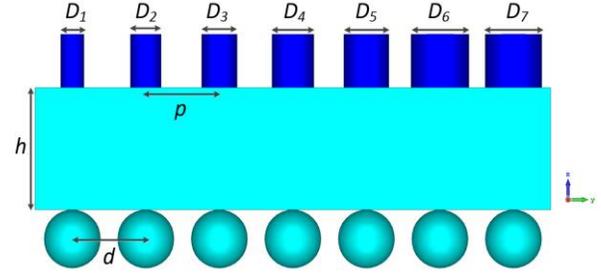

Fig. 9. Unit-cell of the all-dielectric optical reflectarray. The metallic backing layer is replaced by an all-dielectric metasurface working as a reflector.

Using the results in Fig. 10, we can retrieve the diameters of the cylindrical resonators of the reflectarray by choosing the values returning a progressive 60° phase shift between consecutive elements. After a quick optimization process, needed to account for the coupling between the resonators, we obtain the following design parameters: $D_1 = 38$ $nm$, $D_2 = 130$ $nm$, $D_3 = 151$ $nm$, $D_4 = 178$ $nm$, $D_5 = 191$ $nm$, $D_6 = 239$ $nm$, $D_7 = 245$ $nm$. A reflectarray with these parameters has been simulated in a full-wave simulator by imposing periodic boundaries. The structure has been excited from the top with a normally impinging plane wave. In Fig. 11, we report the magnitude of the reflection coefficient of the $S_{41}$ parameter (i.e., the fraction of the impinging power that is coupled with the (01) Floquet mode), the total efficiency of the device and the overall transmission coefficient (i.e., the sum of the transmission coefficients of the impinging wave with all the relevant Floquet modes). Please note that, given the period of the reflectarray $p = 310$ $nm$, the (01) Floquet mode propagates with an angle $\theta = 20°$ with respect the z-axis. As it can be appreciated, at the design frequency $f_0$, most of the impinging power is coupled with this mode. Furthermore, the efficiency of the device reaches 92%. The total transmission is very low, confirming that the designed all-dielectric metasurface works effectively as a reflector. It is also interesting to observe that, notably, the deflection of the impinging wave is achieved within a moderately broad frequency range (i.e., $|S_{41}| > 0.9$ in the range 465-480 THz). In Fig. 12, we show the contour plot of the magnitude electric field at $f_0$ that confirms the expected deflection of the impinging wave.

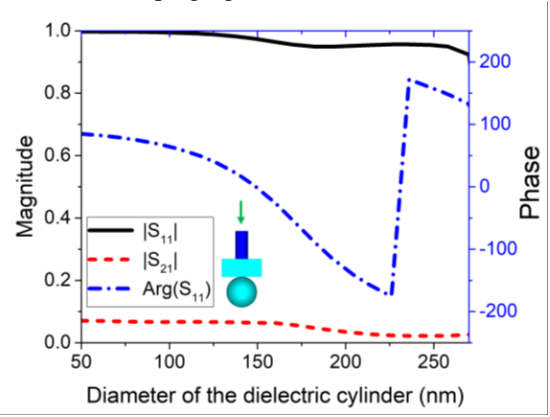

Fig. 10. Reflection (magnitude and phase) and transmission coefficients of the unit-cell of the reflectarray as the diameter of the dielectric cylinder changes.

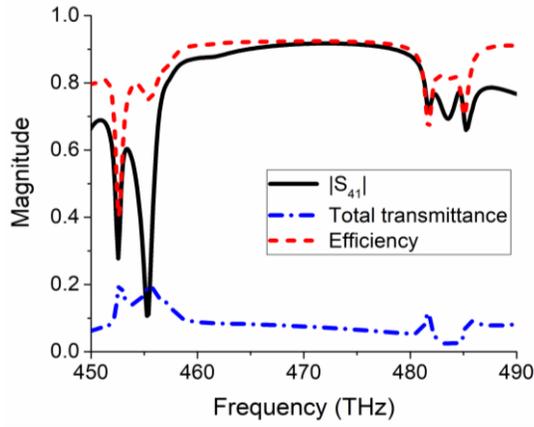

Fig. 11. Magnitude of the $S_{41}$ parameter, total transmittance and efficiency of the designed reflectarray.

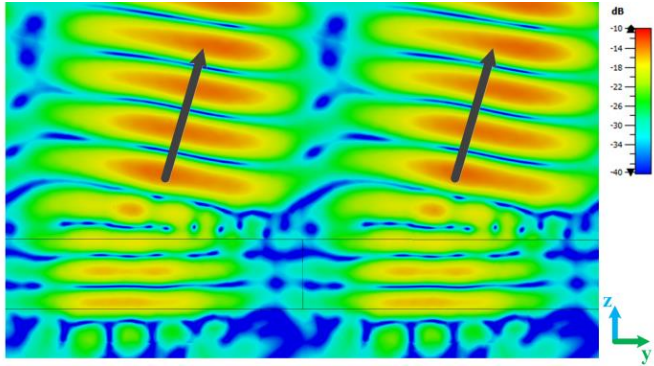

Fig. 12. Contour plot of the magnitude of the electric field at $f_0$. The scale (in *dB*) is normalized to the maximum value.

To the best of our knowledge, this is one of the most efficient optical DRA reflectarray discussed in literature [55] and it is able to overcome the loss limitations affecting conventional optical DRAs discussed in [56]. Differently from the earlier designs [54],[56]-[57], it does not contain plasmonic materials.

*C. Unconventional antenna reflectors*

As a last possible application of all-dielectric metasurfaces, we focus our attention on their use at microwave frequencies and, in particular, for implementing unconventional antenna reflectors. Metallic reflectors are widely used for designing different classes of antennas, including directive dipoles. Here, we show that the use of all-dielectric reflectors may introduce some new degrees of freedom in shaping the antenna radiation pattern.

We consider the case sketched in Fig. 13 of a half-wavelength dipole backed by an all-dielectric microwave reflector. According to the antenna theory, this simple system can be considered as a 2-element array whose radiation pattern is equal to:

$$\mathbf{E}_{tot}(\vartheta,\varphi) = \mathbf{E}_0(\vartheta,\varphi)\left(1 + \Gamma e^{-2ik_0 h \cos\vartheta}\right), \quad (14)$$

where $\mathbf{E}_0(\vartheta,\varphi)$ is the field radiated by the dipole antenna (whose expression can be found in [60]), $\Gamma$ is the complex reflection coefficient of the reflector and *h* is the distance between the antenna and the reflector. For a conventional metallic reflector, $\Gamma$ is always equal to $e^{-i\pi}$ and, thus, the only possibility to shape the radiation pattern of the antenna is to act on the distance *h*. All-dielectric reflectors, conversely, can exhibit different values of the reflection phase for a fixed *h* and, thus, may enable unconventional effects.

To better explain this point, in Fig. 14 we show the reflection magnitude and phase of a microwave reflector consisting of an array of spherical particles made by a material with permittivity $\varepsilon_s = 100$ and $\tan(\delta) = 0.01$. Ceramic materials with similar values of complex dielectric constants are quite common at microwave [58] and even THz frequencies [59], and can be realized with a high degree of homogeneity. The geometrical parameters of the analytically-designed reflector are: $a = 0.5\lambda = 0.05\lambda_0$, $d = 1.2\lambda = 0.12\lambda_0$, being $\lambda$ the wavelength inside the dielectric at $f_0 = 3$ GHz. As it can be appreciated, within the reflection bandwidth of the screen, the reflection phase progressively changes from 0 (i.e., PMC-like behavior) to $\pi$ (i.e., PEC-like behavior). According to (14), we expect that the radiation diagram of the dipole is strongly frequency-dependent. Please note that, compared to the results shown in Fig. 6, the additional narrow reflection peak due to the magnetic quadrupole is here significantly reduced due to the moderate losses of the dielectric.

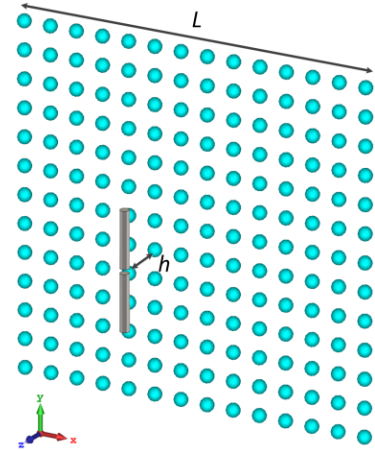

Fig. 13. Half-wavelength dipole backed by an all-dielectric reflector. The transverse dimension of the reflector is $L{\times}L$ ($L=2\lambda_0$).

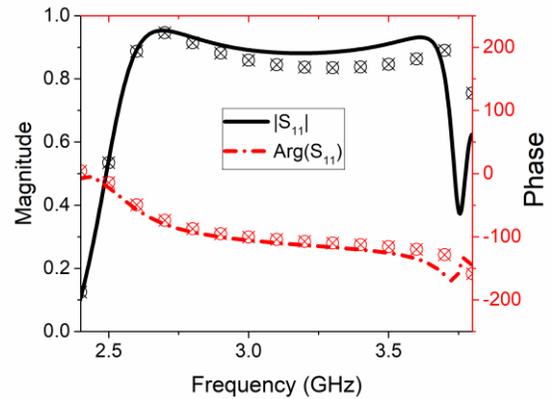

Fig. 14. Magnitude and phase of the reflection coefficient of the designed all-dielectric microwave metasurface. Ticks represent the results of full-wave simulations.

This is confirmed by the results in Fig. 15 where we show the directivity of the system on the horizontal plane, computed through full-wave simulations and assuming $h = \lambda_0/2$. As expected, the main lobe direction of the overall system changes with the frequency due to the variation of the reflection phase of the reflector. The pointing direction of the radiative system covers a wide range of angles on the horizontal plane, i.e., from $\theta = 0°$ to $\theta > 60°$. Specifically, for lower frequencies, where the reflector behaves similarly to a PMC, the dipole has a directive pattern at broadside. Conversely, for higher frequencies, the radiation pattern exhibits a progressively deeper broadside null, due to the PEC-like behavior of the backing screen.

These results are similar to those that could be obtained using an ideal infinitely-extended PEC or PMC reflector. Specifically, for the PMC-case (*i.e.,* 2.6 GHz), we obtain a maximum directivity of 8.96 *dB* and half-power beam-width on the horizontal plane (HPBW-H) equal to 80° (the corresponding results in the ideal case are $D_{max} = 8.65$ *dB* and HPBW-H = 80°). For the PEC-like frequency, instead, the results with the real reflector are: beam direction $\theta_{max} = 60°$, $D_{max} = 6.38$ *dB* and HPBW = 50° (the corresponding results in the ideal case are $\theta_{max} = 61°$, $D_{max} = 8.2$ *dB* and HPBW = 80°). The radiation efficiency of the system is higher than 95 %, due to the reduced Ohmic losses (that are limited to the metallic dipole and to the low losses of the dielectric).

The proposed system may find application in the design of point-to-multipoint radio-links in which the same antenna can be used to communicate, at slightly different frequencies, with other stations distributed around. We would like to stress that, as clarified in Section II, the working principle behind this unconventional reflector is the result of an average interaction between electric and magnetic dipoles excited by the individual spherical particles. This means that, as we have already shown for the case of plasmonic metasurfaces [35], its performance does not critically depend on the exact size and properties of the individual particles. Our numerical simulations (not reported here for the sake of brevity) confirm that the introduction of realistic geometrical and electromagnetic perturbations in the spheres composing the reflector does not affect significantly the frequency response within the dipole bandwidth.

Interestingly, a similar shaping of the radiation pattern can be also achieved at a given frequency by acting on the permittivity $\varepsilon_h$ of the host medium. If the material hosting the particles is made tunable, for example by using liquid crystals or plasmas, it would become possible tuning electronically the reflection phase and, thus, the pointing direction and the null depth of the radiative system without acting on the antenna excitation. The shaping of the radiation patterns at a fixed frequency may find applications in the design of switched beam and smart antennas for interference rejection and electronic defense.

Before concluding, we underline that the possibility to tune the reflection phase of an antenna reflector has been already discussed for metallic metasurfaces (see, for instance, [61]) or magnetodielectric particles [62] but here is described relying on low-loss and purely dielectric materials. This paves the way to high-power applications of unconventional reflectors. Among the others, we remember that artificial magnetic conductors find application in RCS reduction [63] and antenna miniaturization [64].

## V. CONCLUSIONS

In this paper, we have discussed a simple and reliable homogenization technique for describing the electromagnetic behavior of all-dielectric metasurfaces. The model is based on a combination of the Mie theory for spherical particles with a surface-impedance homogenization and it succeeds in fully characterize the response of a metasurface supporting electric and magnetic induced dipole moments. The proposed approach allows deriving closed-form expressions of the effective electric and magnetic surface impedances of an all-dielectric metasurface and of its reflection and transmission coefficient. Full-wave simulations, carried out in different scenarios, confirm that the proposed technique is effective even for very low inter-element separation distances. In addition, we have shown how the surface impedance model of all-dielectric metasurfaces may be used in different applicative scenarios, spanning from the design of optical reflectarrays up to the design of frequency-dispersive reflectors for microwave antennas. The availability of a simple and effective model for all-dielectric metasurfaces may boost their use in realistic applications, including nanophotonics components and high-power microwave devices.

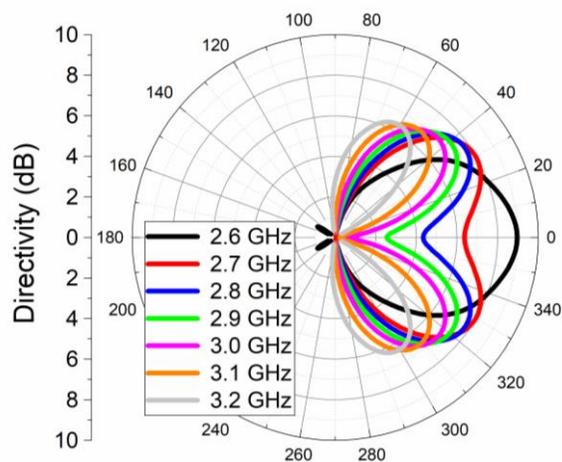

Fig. 15. Full-wave directivity of the system shown in Fig. 12 on the plane $\varphi = 0°$ at different frequencies within the reflection bandwidth of the reflector.

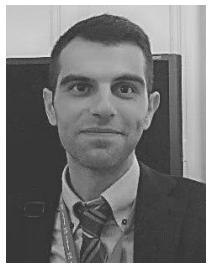

**Alessio Monti** (S'12–M'15–SM'19) received the B.S. degree (summa cum laude) and the M.S. degree (summa cum laude) in electronic and ICT engineering both from ROMA TRE University, Rome, in 2008 and 2010, respectively. From 2011 to 2013, he attended the doctoral school in biomedical electronics, electromagnetics and telecommunications engineering at ROMA TRE University. Currently, he is an assistant professor at Niccolò Cusano University, Rome, Italy, where he teaches courses on antenna and microwave theory.

Dr. Monti is member of the secretariat office of the International Association METAMORPHOSE VI, and of the Editorial Board of the journals *IEEE Transaction on Antennas and Propagation* and *EPJ Applied Metamaterials*. In 2019, he has served as General Chair of the *International Congress on Artificial Materials for Novel Wave Phenomena – Metamaterials* and has been serving as Chair of the Steering Committee of the same Congress series since 2017. He is also member of the Technical Program Committee (TPC) of the *IEEE International Symposium on Antennas and Propagation* since 2016 and has been member of the TPC of the *International Congress on Advanced Electromagnetic Materials in Microwaves and Optics- Metamaterials* during the years 2014-2016. He has also been serving as a Technical Reviewer of many high-level international journals related to electromagnetic field theory, metamaterials and plasmonics and he been selected as one of the Top Reviewers by the Editorial Board of the *IEEE Transactions on Antennas & Propagation* for several years.

His research interests include the design and the applications of microwave and optical artificially engineered materials and metasurfaces, the design of cloaking devices for scattering cancellation at microwave and optical frequencies with a particular emphasis on their applications to the antenna theory. His research activities resulted in more than 80 papers published in international journals and conference proceedings. Dr. Monti has been the recipient of some national and international awards and recognitions, including the *URSI Young Scientist Award* (2019), the *outstanding Associate Editor* of the *IEEE Transactions on Antennas and Propagation* (2019), the *Finmeccanica Group Innovation Award for young people* (2015) and the 2nd place at the student paper competition of the conference *Metamaterials'* (2012).

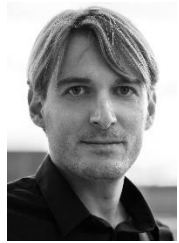

**Andrea Alù** (S'03, M'07, SM'12, F'14) is the Founding Director of the Photonics Initiative at the Advanced Science Research Center (ASRC) at the Graduate Center of the City University of New York (CUNY). He is also the Einstein Professor of Physics at the CUNY Graduate Center and Professor of Electrical Engineering at the City College of New York, and an Adjunct Professor and Senior Research Scientist at the University of Texas at Austin. He received the *Laurea*, MS and PhD degrees from the University of Roma Tre, Rome, Italy, respectively in 2001, 2003 and 2007. From 2002 to 2008, he has been periodically working at the University of Pennsylvania, Philadelphia, PA, where he has also developed significant parts of his PhD and postgraduate research. After spending one year as a postdoctoral research fellow at UPenn, in 2009 he joined the faculty of the University of Texas at Austin, where he was the Temple Foundation Endowed Professor until 2018.

He is the co-author of an edited book on optical antennas, over 450 journal papers and over 30 book chapters, with over 30,000 citations to date. He has organized and chaired various special sessions in international symposia and conferences, and was the technical program chair for the IEEE AP-S symposium in 2016, and for the international Metamaterials conference in 2014 and 2015. His current research interests span over a broad range of areas, including metamaterials and plasmonics, electromagnetics, optics and nanophotonics, acoustics, scattering, nanocircuits and nanostructures, miniaturized antennas and nanoantennas, RF antennas and circuits.

Dr. Alù is currently an Associate Editor of *Applied Physics Letters* and serves on the Editorial Board of *Physical Review B*, *Advanced Optical Materials*, *EPJ Applied Metamaterials* and *ISTE Metamaterials*. He served as Associate Editor for the *New Journal of Physics*, *MDPI Materials*, *IEEE Antennas and Wireless Propagation Letters*, *Scientific Reports*, *Metamaterials*, *Advanced Electromagnetics*, and *Optics Express*. He has guest edited special issues for the *IEEE Journal of Selected Topics in Quantum Electronics*, *Proceedings of IEEE*, *IEEE Transactions on Antennas and Propagation*, *IEEE Antennas and Wireless Propagation Letters*, *Nanophotonics*, *Journal of Optics*, Journal of the Optical Society of America B, *Photonics and Nanostructures - Fundamentals and Applications*, *Optics Communications*, *Metamaterials*, and *Sensors* on a variety of topics involving metamaterials, plasmonics, optics and electromagnetic theory.

Over the last few years, he has received several research awards, including the *IEEE Kiyo Tomiyasu Award* (2019), *IUMRS Young Researcher Award* (2018), *ICO Prize in Optics* (2016), the inaugural *MDPI Materials Young Investigator Award* (2016), the *Kavli Foundation Early Career Lectureship in Materials Science* (2016), the inaugural *ACS Photonics Young Investigator Award Lectureship* (2016), the *Edith and Peter O'Donnell Award in Engineering* (2016), the *NSF Alan T. Waterman Award* (2015), the *IEEE MTT Outstanding Young Engineer Award* (2014), the *OSA Adolph Lomb Medal* (2013), the *IUPAP Young Scientist Prize in Optics* (2013), the *Franco Strazzabosco Award for Young Engineers* (2013), the *SPIE Early Career Investigator Award* (2012), the *URSI Issac Koga Gold Medal* (2011), an *NSF CAREER award* (2010), the *AFOSR* and the *DTRA Young Investigator Awards* (2010, 2011), *Young Scientist Awards* from *URSI General Assembly* (2005) and *URSI Commission B* (2010, 2007 and 2004). His students have also received several awards, including student paper awards at *IEEE Antennas and Propagation Symposia* (in 2011 to Y. Zhao, in 2012 to J. Soric). He is a *Simons Investigator in Physics* since 2016, has been selected four times as finalist of the *Blavatnik Award for Young Scientists* (2016, 2017, 2018, 2019), and of the *IET A F Harvey Prize Engineering Research Prize* in 2019, and he has been a *Highly Cited Researcher* from *Web of Science* since 2017. He has been serving as the President of the *Metamorphose Virtual Institute for Artificial Electromagnetic Materials and Metamaterials*, as a member of the Administrative Committee of the *IEEE Antennas and Propagation Society*, as an *OSA Traveling Lecturer* since 2010, as an *IEEE AP-S Distinguished Lecturer* since 2014, and as the *IEEE joint AP-S and MTT-S chapter for Central Texas*. Finally, he is a full member of *URSI*, a Fellow of *OSA*, *AAAS*, *SPIE* and *APS*.

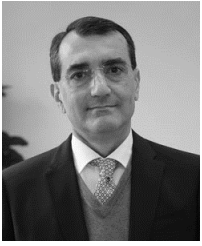

**Alessandro Toscano** (M'91-SM'11) (Capua, 1964) graduated in Electronic Engineering from Sapienza University of Rome in 1988 and he received his PhD in 1993. Since 2011, he has been Full Professor of Electromagnetic Fields at the Engineering Department of ROMA TRE University. He carries out an intense academic and scientific activity, both nationally and internationally.

From April 2013 to January 2018 he was a member of ROMA TRE University *Academic Senate*. From October 2016 to October 2018, he is a member of the National Commission which enables National Scientific Qualifications to Full and Associate Professors in the tender sector 09/F1 – Electromagnetic fields.

Since 23rd January 2018 he has been *Vice-Rector* for Innovation and Technology Transfer.

In addition to his commitment in organizing scientific events, he also carries out an intense editorial activity as a member of the review committees of major international journals and conferences in the field of applied electromagnetics. He has held numerous invited lectures at universities, public and private research institutions, national and international companies on the subject of artificial electromagnetic materials, metamaterials and their applications. He actively participated in founding the international association on metamaterials *Virtual Institute for Advanced Electromagnetic Materials – METAMORPHOSE, VI*. He coordinates and participates in several research projects and contracts funded by national and international public and private research institutions and industries.

Alessandro Toscano's scientific research has as ultimate objective the conceiving, designing and manufacturing of innovative electromagnetic components with a high technological content that show enhanced performance compared to those obtained with traditional technologies and that respond to the need for environment and human health protection. His research activities are focused on three fields: metamaterials and unconventional materials, in collaboration with Professor A. Alù's group at The University of Texas at Austin, USA, research and development of electromagnetic cloaking devices and their applications (First place winner of the *Leonardo Group Innovation Award* for the research project entitled: 'Metamaterials and electromagnetic invisibility') and the research and manufacturing of innovative antenna systems and miniaturized components (first place winner of the *Leonardo Group Innovation Award* for the research project entitled: "Use of metamaterials for miniaturization of components" – MiniMETRIS).

He is the author of more than one hundred publications in international journals indexed ISI or Scopus; of these on a worldwide scale, three are in the first 0.1 percentile, five in the first 1 percentile and twenty-five in the first 5 percentile in terms of number of quotations and journal quality.

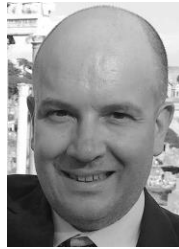

**Filiberto Bilotti** (S'97–M'02–SM'06–F'17) received the Laurea and Ph.D. degrees in electronic engineering from ROMA TRE University, Rome, Italy, in 1998 and 2002, respectively. Since 2002, he has been with the Faculty of Engineering (2002-2012) and, then, with the Department of Engineering (2013-now), at ROMA TRE University, where he serves as a Full Professor of electromagnetic field theory (2014-now) and the *Director of the Antennas and Metamaterials Research Laboratory* (2012-present)..

His main research contributions are in the analysis and design of microwave antennas and arrays, analytical modelling of artificial electromagnetic materials, metamaterials, and metasurfaces, including their applications at both microwave and optical frequencies. In the last ten years, Filiberto Bilotti's main research interests have been focused on the analysis and design of cloaking metasurfaces for antenna systems, on the modelling and applications of (space and) time-varying metasurfaces, on the topological-based design of antennas supporting structured field, on the modelling, design, and implementation of non-linear and reconfigurable metasurfaces, on the concept of meta-gratings and related applications in optics and at microwaves, on the modelling and applications of optical metasurfaces. The research activities developed in the last 20 years (1999-2019) has resulted in more than 500 papers in international journals, conference proceedings, book chapters, and 3 patents.

Prof. Bilotti has been serving the scientific community, by playing leading roles in the management of scientific societies, in the editorial board of international journals, and in the organization of conferences and courses.

In particular, he was a founding member of the *Virtual Institute for Artificial Electromagnetic Materials and Metamaterials – METAMORPHOSE VI* in 2007. He was elected as a member of the Board of Directors of the same society for two terms (2007-2013) and as the President for two terms (2013-2019). Currently, he serves the METAMORPHOSE VI as the Vice President and the Executive Director (2019-now).

Filiberto Bilotti served as an Associate Editor for the *IEEE Transactions on Antennas and Propagation* (2013-2017) and the journal *Metamaterials* (2007-2013) and as a member of the Editorial Board of the *International Journal on RF and Microwave Computer-Aided Engineering* (2009-2015), *Nature Scientific Reports* (2013-2016), and *EPJ Applied Metamaterials* (2013-now). He was also the Guest Editor of 5 special issues in international journals.

He hosted in 2007 the inaugural edition of the *International Congress on Advanced Electromagnetic Materials in Microwaves and Optics – Metamaterials Congress*, served as the Chair of the Steering Committee of the same conference for 8 editions (2008-2014, 2019), and was elected as the General Chair of the *Metamaterials Congress* for the period 2015-2018. Filiberto Bilotti was also the General Chair of the *Second International Workshop on Metamaterials-by-Design Theory, Methods, and Applications to Communications and Sensing* (2016) and has been serving as the chair or a member of the technical program, steering, and organizing committee of the main national and international conferences in the field of applied electromagnetics.

Prof. Bilotti was the recipient of a number of awards and recognitions, including the elevation to the *IEEE Fellow* grade for contributions to metamaterials for electromagnetic and antenna applications (2017), *outstanding Associate Editor of the IEEE Transactions on Antennas and Propagation* (2016), *NATO SET Panel Excellence Award* (2016), *Finmeccanica Group Innovation Prize* (2014), *Finmeccanica Corporate Innovation Prize* (2014), IET Best Poster Paper Award (Metamaterials 2013 and Metamaterials 2011), Raj Mittra Travel Grant Senior Researcher Award (2007).